\documentstyle[11pt,paspconf,epsfig]{article}
\begin{document}

\title{Population Synthesis of the GRB Progenitors and their
       Brightness and Redshift Distribution 
}

\author{I. Panchenko\altaffilmark{1}}
\affil{Department of Physics, Moscow State University,
    Moscow, Russia}
\affil{Sternberg Astronomical Institute, Moscow, Russia}
\altaffiltext{1}{ivan@sai.msu.su}

\def\lns{ $\log N$-$\log P$ }
\def\P{Paczy\'nski }
\def\J{J{\o}rgensen }

\begin{abstract}

We estimate the relativistic binaries merger rate and  redshift distribution 
population synthesis of an ensemble of evolving close binaries. 
Results of such simulations definitely depend on the
cosmic star formation rate history, which is  different in galaxies
of different types. This leads to a difference in merger rate in spiral and
elliptical galaxies, which in principle can be an observational test for
GRB models.
Also a fit of BATSE long bursts \lns is performed, showing that 
only a wide ($2-3$ orders of magnitude) luminosity function 
provides a good fit.
\end{abstract}


\section{Introduction}
For many years the binary relativistic star merger remains one of the most 
valuable models of gamma ray bursts (Blinnikov et al, 1984, \P, 1991).
Some recent alternative models, involving a collapse of massive rotating star
(\P, 1998), possibly easier fit the energy requirements for GRB,
and are in nice agreement with the discoveries of GRB optical counterparts in 
actively star forming galaxies (Bloom et al, 1998).

Nevertheless, the old merger model is not excluded, 
as the observed binary radiopulsars display the examples of
definite merger precursors.

This paper is devoted to the studies of close binary evolution which leads 
to the mergers of binary neutron star and black hole systems (NS+NS and NS+BH)
which are suspected to be able to produce a GRB.

The collapsar GRB and the merger GRB can have 
similar physical mechanisms for energy extraction (Lee, 1999)
and production of radiation (Piran, 1999, Usov, 1999).
Nevertheless, they can be discriminated  by evolutionary considerations.

\section{The model of evolution}

We use the ``Scenario Machine'', initiated by Kornilov and Lipunov (1995), 
and later developed by Lipunov et al (1996), as a stellar evolution engine.
The basic evolution model is similar to one of  Vanbeveren et al (1998), 
which are the best to reproduce  the galactic population of massive binaries.

During the evolution of a binary its orbital separation changes in various 
modes of mass and angular momentum transfer. The most dramatic change of
the orbit takes place during a supernova explosion, or a collapse into a 
black hole. It is usually supposed that a newly formed compact object 
obtains a kick velocity due to some asymmetry in the collapse. To explain the
observed velocities of radio pulsars, one should assume the kick velocity 
to be of the order of $200-600$~km/s (Lyne, Lorimer, 1994).

As the observations suffer from many selection effects, the 
exact shape of the distribution is not precisely known. So we perform the
calculations for several values of the kick velocity, in the range from $0$
to $600$~km/s, to show the influence of this parameter on the results.

\section{The event age distribution}

The life of the merging system consists of two important phases: first, the 
nuclear powered evolution of the normal stars, and second, the gravity wave powered
orbit shrinking phase. The characteristic time scale of the first phase occupies the range 
from $3\cdot10^6$ years for the most massive stars to $\sim 10^8$years for the 
least massive stars being able to produce a neutron star.
The gravitational inspiral time is determined by the parameters of the orbit 
after the formation of second compact object, and can be much greater
than nuclear lifetime, being on average of the order of a billion years.

\begin{figure}
 \centerline{\epsfig{file=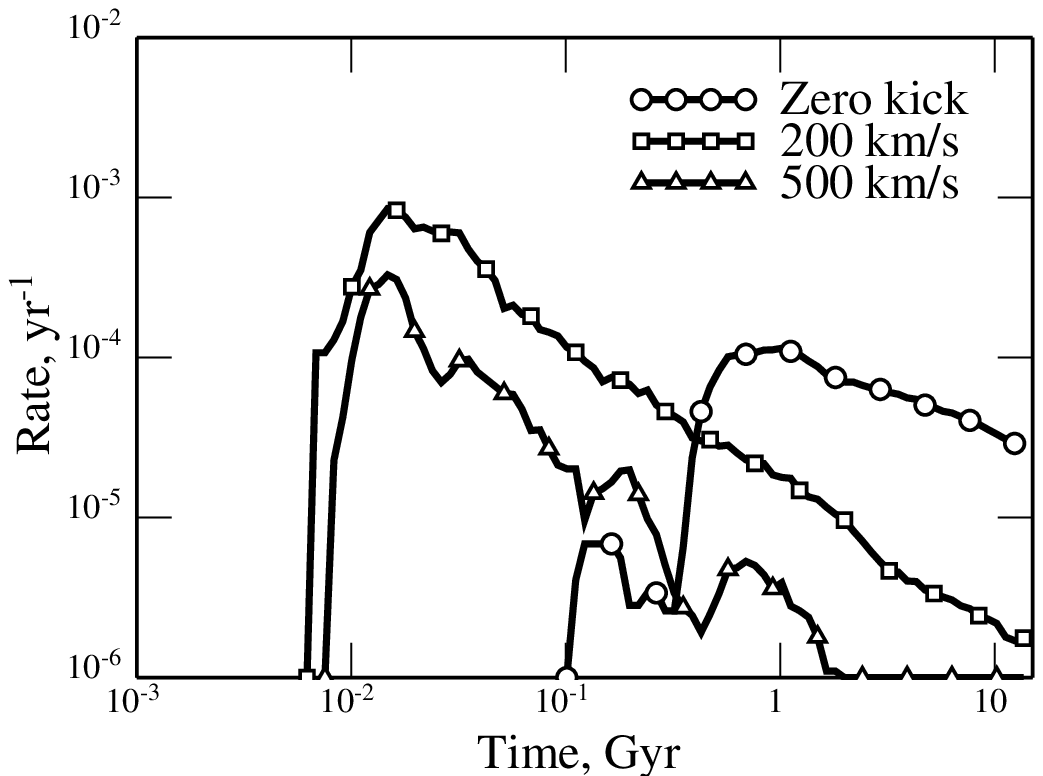,width=7cm}
             \epsfig{file=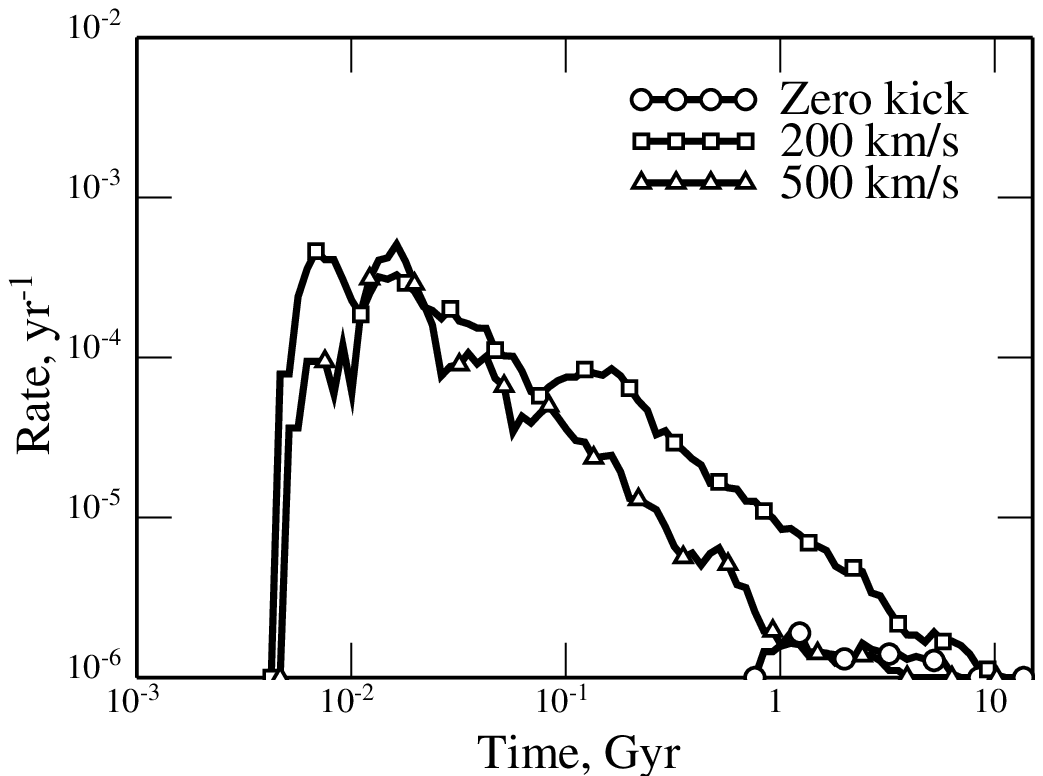,width=7cm}
 }
 \caption{The binary merger age distribution. Shown as the event rate after 
 a simultaneous formation of a typical galaxy ($10^{11} M_\odot$). Left: NS+NS,
 Right: NS+BH.
 }
 \label{greens}
\end{figure}

The age of a merging binary is a sum of the nuclear lifetime and 
the gravitational inspiral duration, 
The distribution of the ages of merging binaries is a direct output of the
population synthesis procedure. Another  sense of this distribution 
is the merger rate time history after a simultaneous ($\delta$-like) 
star formation. It is shown in fig.\ref{greens} for both NS+NS and NS+BH mergers.

The merger age distribution shows a power-law behavior with a slope 
$\approx -1$. For NS+NS mergers, there is a strong dependency of the 
sharpness of the age distribution on the kick velocity. High kicks make 
the mergers on average younger. Without a kick, there are no mergers with
ages less than $10^8$~years. 

For NS+BH mergers, the presence of a kick velocity is very important: without 
a kick, all the binaries are very wide and have very long inspiral times, 
usually greater than the Hubble time.

The power law distribution has a ``heavy tail'': though a half of the mergers
take place before several hundred million years after the star formation, 
there is also a significant part of  mergers that take part after billions 
of years.

\section{The distribution of merger redshifts}

The distribution of merger redshifts can be obtained by a convolution of the 
star formation rate history and the merger age distribution. The present 
knowledge of the star formation rate (SFR) history is based on the works of 
Madau (1996). Earlier (\J et al, 1995) we assumed a simple two-parametric
model of SFR, containing an initial star formation burst (where $\epsilon$
of all the stars were formed) at $z\sim 5$, and a subsequent constant SFR.

In this work we base on the Madau observational SFR, but still introduce
an initial star formation burst, which should be responsible for the formation of the 
elliptical galaxies and spheroidal components of the spiral galaxies. 
Then, according to Fukugita et al (1998), the value of $\epsilon$ should be 
$\sim 2/3$.
The adopted cosmological model was 
($H_0=75$ km/s/Mpc, $\Omega_m=0.3$, $\Omega_\Lambda=0.7$).

\begin{figure}
 \centerline{\epsfig{file=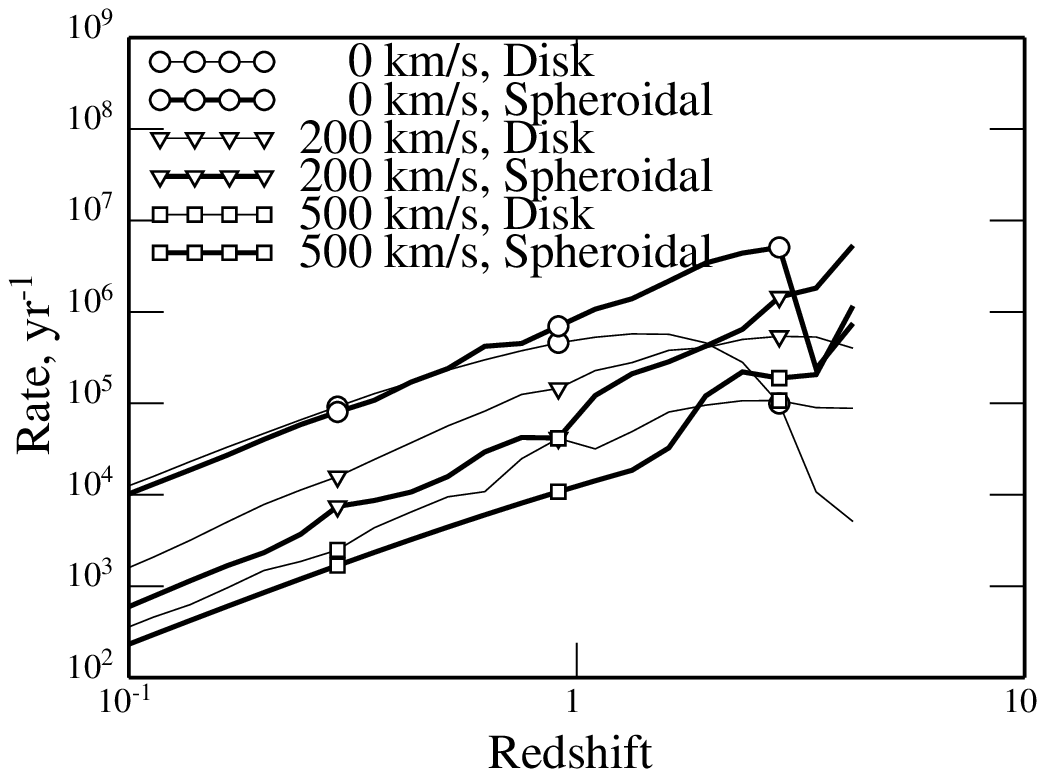,width=7cm}
             \epsfig{file=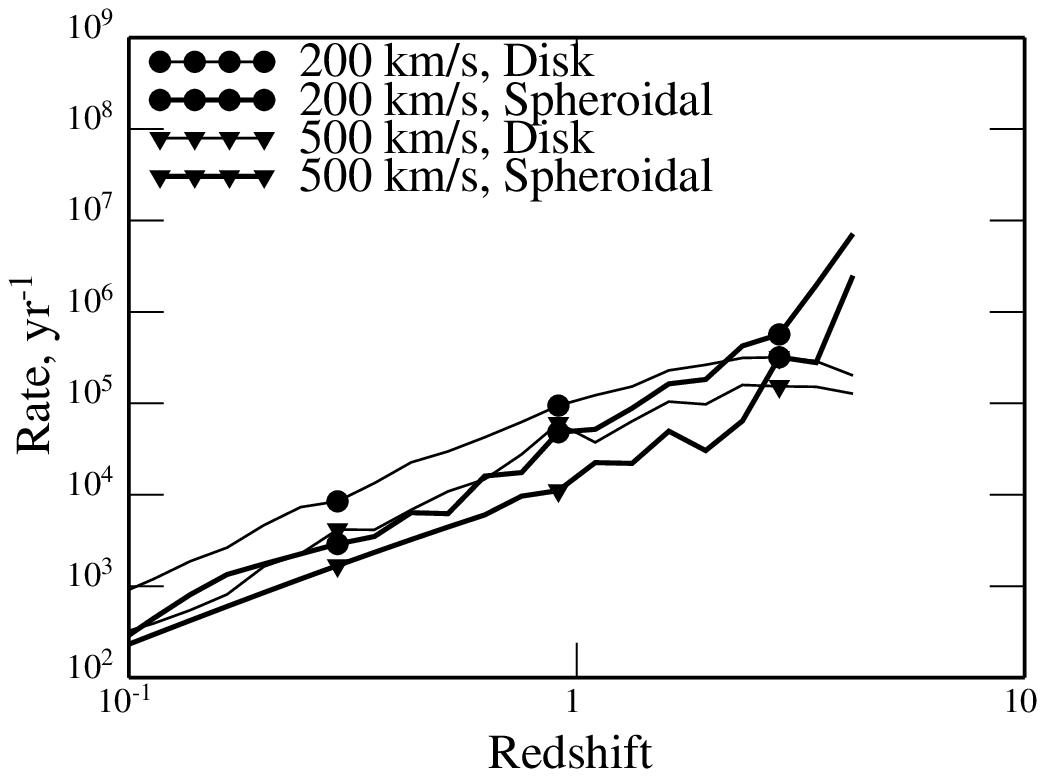,width=7cm}}
 \caption{The merger redshift distributions (rate per unit $z$). 
          Left: NS+NS; Right: NS+BH. }
 \label{z_nnb}
\end{figure}

The obtained NS+NS and NS+BH merger redshift distributions  are displayed in 
fig.~\ref{z_nnb} for spiral and elliptical galaxies (Madau and burst-like SFR,
respectively) and different values
of mean kick velocity. 

Obviously, these event redshift distributions are different for spirals 
and ellipticals. This allows to propose a new observational test for 
the GRB progenitor. If we detect a GRB in an elliptical galaxy, it should 
be most likely a NS+NS and NS+BH merger and not a collapsar, because the 
population of the ellipticals is very old and does not contain massive normal 
stars. 

Thus, future identifications of GRB host galaxies can solve the collapsar/merger
dilemma.

\begin{figure}
\centerline{\epsfig{file=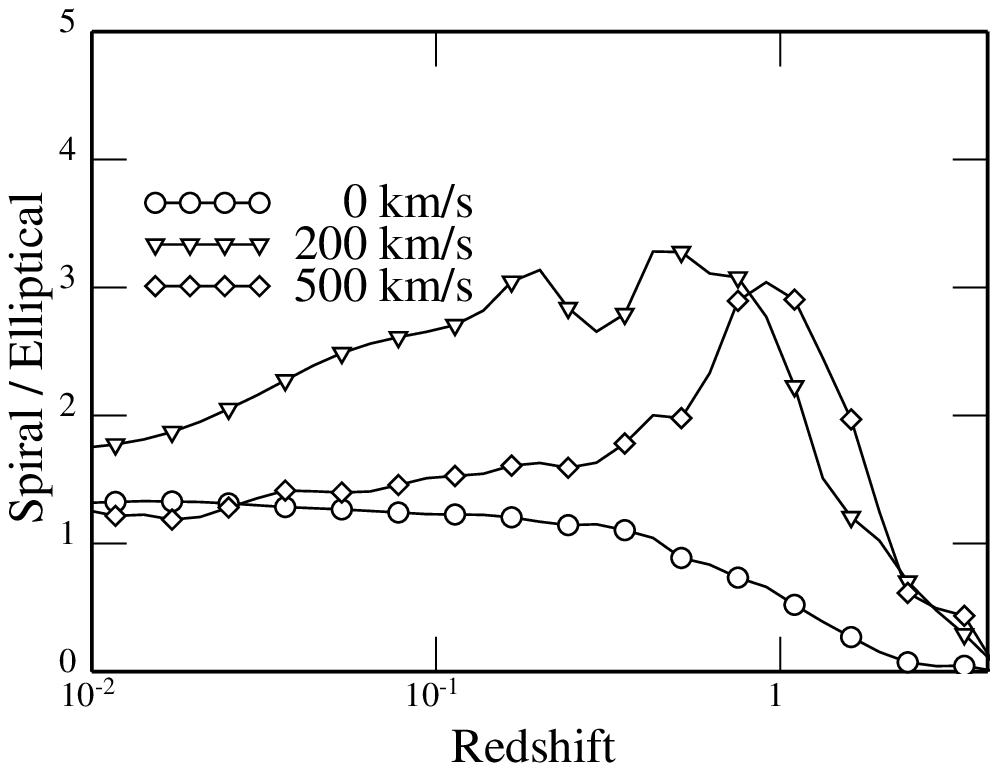,width=7cm}
            \epsfig{file=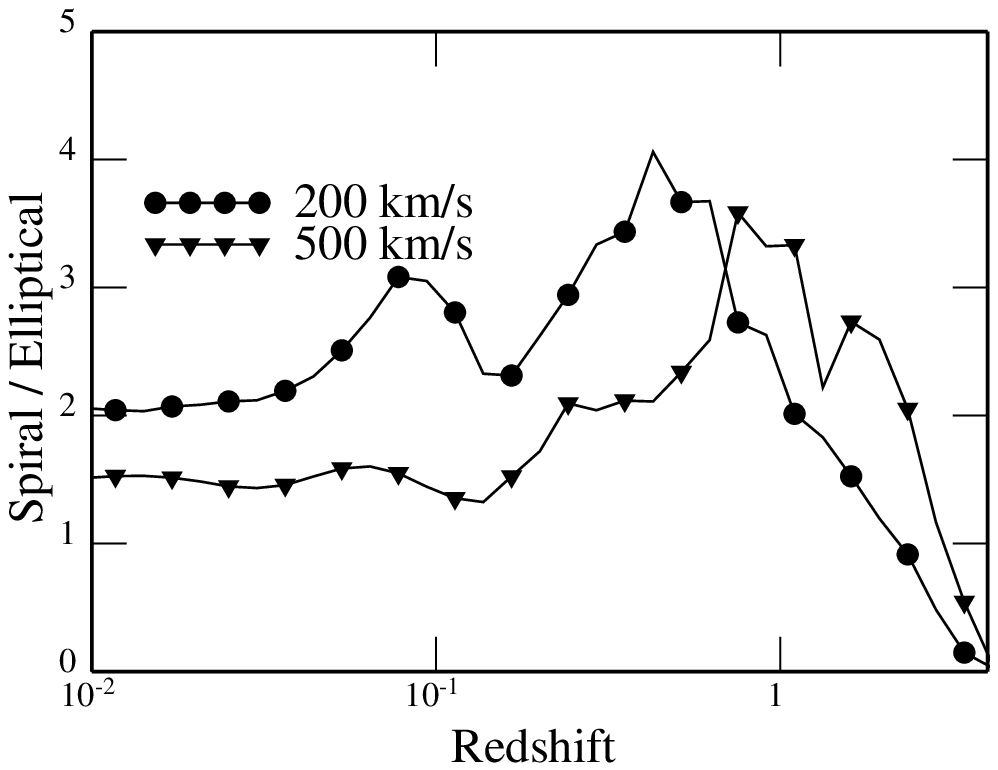,width=7cm}}
 \caption{The evolution of the ratio of merger rates in spiral and elliptical 
          galaxies. Left: NS+NS; Right: NS+BH. }
 \label{zr_nnb}
\end{figure}

\section{The GRB brightness distribution -- \lns }
Another implication of the merger age distributions  is an attempt to 
construct a GRB brightness distribution (\lns). This is done by 
a classic procedure (Weinberg, 1972) on the basis of the redshift distribution
(fig.~\ref{z_nnb}) and a luminosity function. Observations do not 
evidence directly for the spread of the GRB intrinsic luminosities, 
though some techniques allow to estimate it roughly (Petrosian, 1999).
We assumed a power-law luminosity function characterized by the slope and 
range. 

For comparison with the observations, we have chosen the long
($T_{90}>1.5$s) and relatively bright ($>1$~photon/cm$^2$) bursts from 
4th BATSE GRB catalog (Meegan  et al, 1998).
Realistic GRB spectra were taken to estimate the spectral K-correction
(Oke, Sandage, 1968).

\begin{figure}
\centerline{\epsfig{file=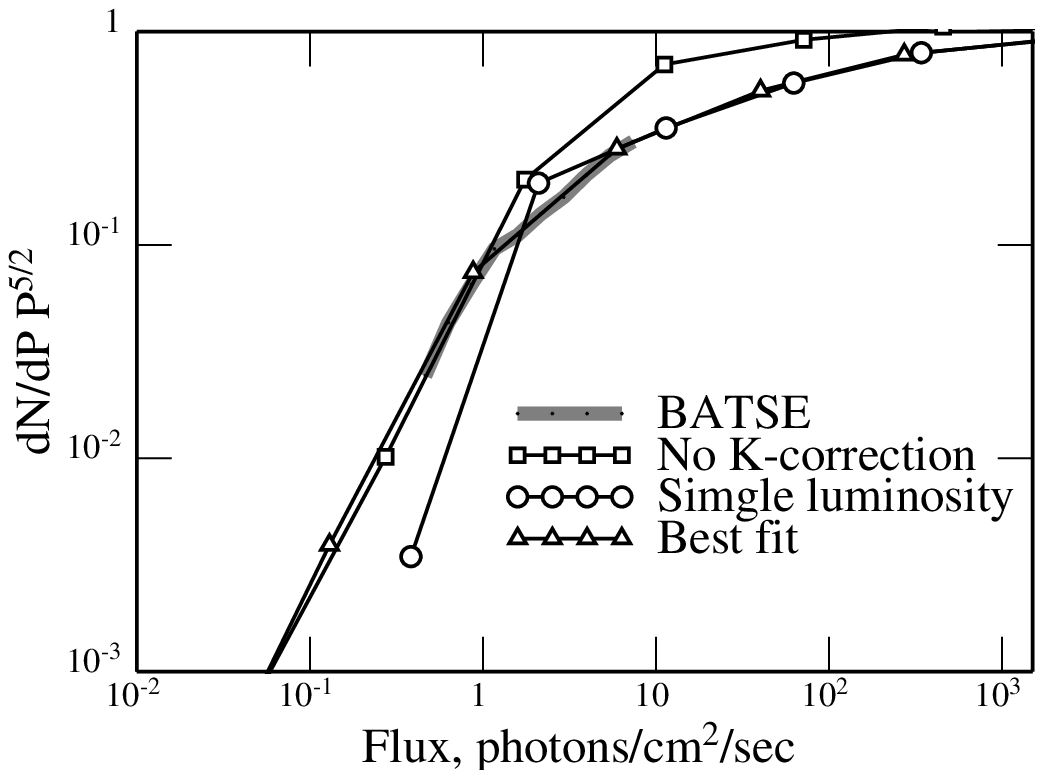,width=7cm}
            \epsfig{file=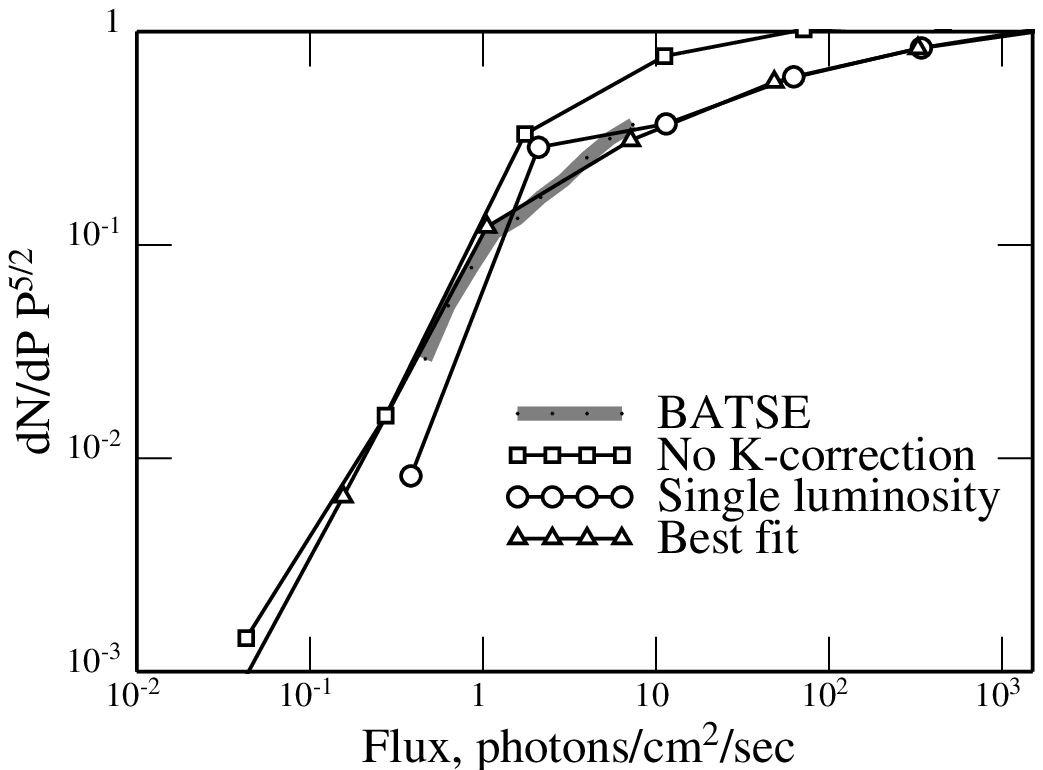,width=7cm}}
\caption{The best fits of the BATSE long bursts \lns.
Left: NS+NS; Right: NS+BH.
Triangles: The best fit. Circles: Without a luminosity spread. Boxes: Without the 
spectral K-correction. Thick line represents the BATSE observations. 
}
\label{lns}
\end{figure}

The best fits are displayed in fig.~\ref{lns}. The \lns distribution is 
shown in a differential form multiplied by $P^{5/2}$ in order to outline
 the difference from the euclidean case, in which the distribution should look 
as a horizontal line.

No acceptable fit can be obtained if luminosity spread is less than $2$ orders
of magnitude. In the best fit, the luminosity function spread is $\approx 2.5$
orders of magnitude, and the most distant observed GRB are at redshift 
$\approx 4.5$. 

\section{Conclusions}

The computer simulations of the evolution of the possible GRB progenitors
provide useful information which facilitates interpretation of the GRB statistics.
In the present paper, we would like to outline two main conclusions:

A new interesting method which can help to determine the nature of 
GRB progenitors is
the determination of the ratio of the rates of bursts in spiral and 
elliptical hosts. For GRB produced by mergers, the rate in ellipticals
should be only several times less than one in spirals, and increase with the redshift.
For the massive collapsar GRB it is practically impossible to occur in an elliptical
galaxy.

The analysis of the observed BATSE \lns distribution with taking into account
the evolutionary effects shows that the the spread of GRB intrinsic luminosity function 
can not be less than two orders of magnitude.

\acknowledgments

Author is grateful to the conference organizers,  Dr Roland
Svensson and  Dr Juri Poutannen for their hospitality.
This work was performed under supported of the 
INTAS project No. 96-0315, the ``Universities of Russia'' project No. 5559 and
RFBR grant No. 98-02-16801.

\end{document}